\documentclass[showpacs,preprintnumbers,apl,superscriptaddress]{revtex4}
\usepackage{graphicx}

\begin{document}

\newcommand {\be}{\begin{equation}}
\newcommand {\ee}{\end{equation}}
\newcommand {\bea}{\begin{eqnarray}}
\newcommand {\eea}{\end{eqnarray}}
\newcommand {\nn}{\nonumber}

\title{Adaptive Design of
Excitonic  Absorption in Broken-Symmetry Quantum Wells}
\author{Jason Thalken}
\affiliation{Department of Physics and Astronomy, University of Southern California, Los
Angeles, CA 90089-0484}
\author{Weifei Li}
\affiliation{Department of Physics and Astronomy, University of Southern California, Los
Angeles, CA 90089-0484}
\author{Stephan Haas}
\affiliation{Department of Physics and Astronomy, University of Southern California, Los
Angeles, CA 90089-0484}
\author{A.F.J. Levi}
\affiliation{Department of Physics and Astronomy, University of Southern California, Los
Angeles, CA 90089-0484}
\affiliation{Department of Electrical Engineering, University of Southern California, Los
Angeles, CA 90089-2533}
\date{\today}
\pacs{73.22.-f,73.22.Gk,61.46.+w,74.78.Na}

\begin{abstract}
Adaptive quantum design is used to identify broken-symmetry
quantum well potential profiles with optical response properties
superior to previous ad-hoc solutions. This technique performs an
unbiased stochastic search of configuration space. It allows us to
engineer many-body excitonic wave functions and thus provides a new
methodology to efficiently develop optimized quantum confined Stark
effect device structures.
\end{abstract}

\maketitle

Excitonic optical absorption at near band gap photon energies in III-V
compound semiconductor quantum well structures is of great
interest for device
applications.  By applying an electric field perpendicular to the plane
of the quantum well, the excitonic optical absorption strength and energy
can be manipulated.  This ``quantum confined Stark effect" (QCSE)\cite{miller}
requires
that electron confinement by the quantum well potential influences electric
field dependent absorption.  Compared to bulk semiconductors, the excitonic
absorption strength in QCSE structures is greater, even in the presence of
large externally applied electric field.  This performance advantage is the
reason why the QCSE has
been used to design novel optical modulators and
detectors.\cite{miller,miller2,huang}
Typically, such designs make use of simple rectangular
potential wells in the AlGaAs/GaAs or InP/InGaAsP material system.  However,
conventional
ad-hoc approaches to device design do not fully exploit the ability of
modern crystal growth techniques to vary the quantum well potential profile
on an atomic monolayer scale in the growth direction.

In this letter, we introduce an adaptive quantum design methodology
that can be used to find a desired target response that is best
suited for a QCSE device.  In contrast to the conventional
approach, we perform an unbiased search of design space to find
the quantum well potential profile $V(x)$ that most closely
approaches the target response.  Because the model of exciton
absorption is a many-body effect, the adaptive quantum design
algorithm may be thought of as manipulating a many-body wave
function to achieve a desired behavior by varying the potential
profile, $V(x)$.

\begin{figure}[h]
\includegraphics[width=8.5cm]{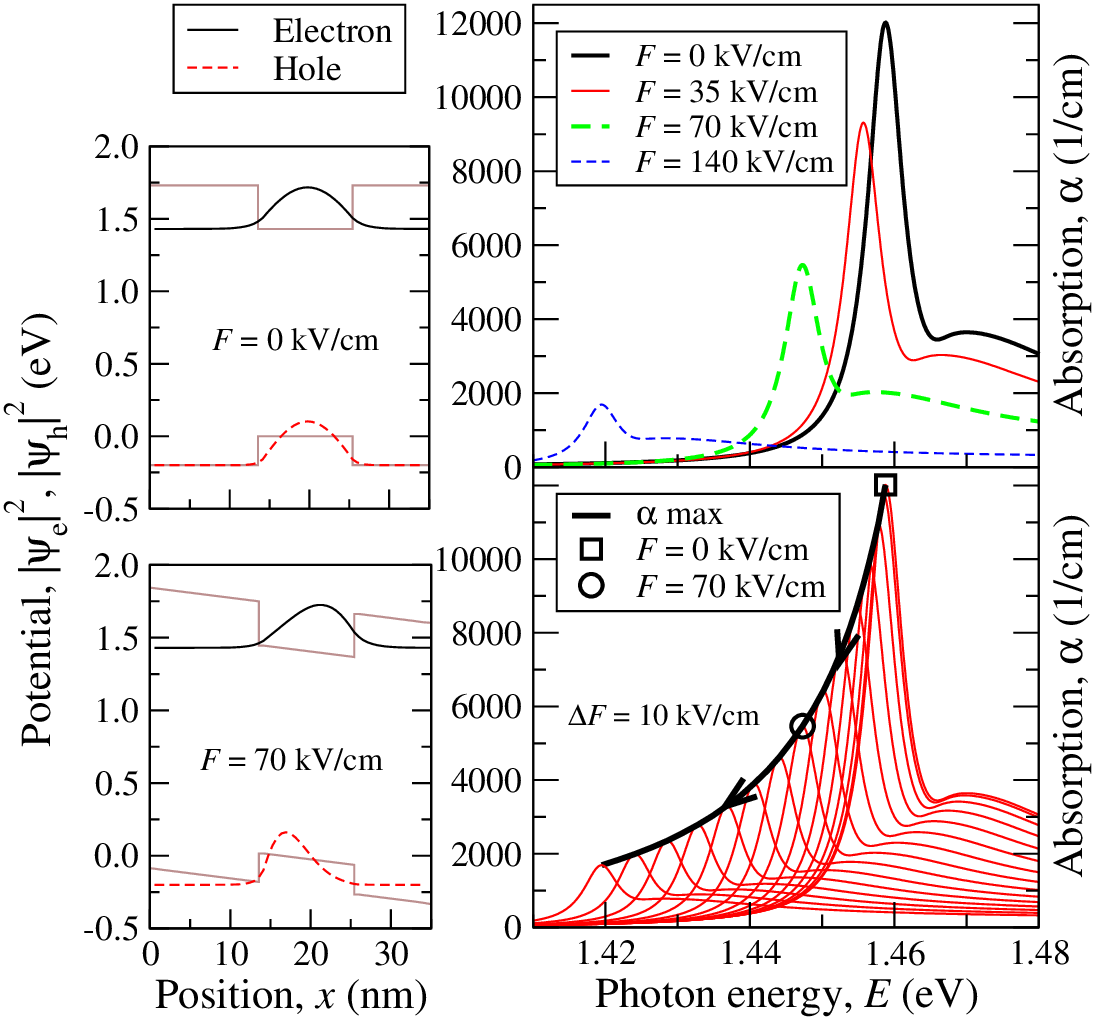}
\caption{Absorption spectrum of a rectangular quantum
well of width 10 nm. In the left panel, the electron (solid curve)
and hole (broken curve) wave functions are shown along with the
profile of the well potential. At finite electric field, $ F $,
applied in the $x$-direction the confining potential is tilted,
shifting the wave functions and reducing their spatial overlap. In
the right panel, the zero-temperature absorption spectrum is shown
for various bias voltages. The QCSE leads to a shift of the
dominant exciton peak toward lower energies, and to a strong
reduction of the maximum absorption with incremental increase in
field, $\Delta F $. }
\end{figure}

Here, we model the QCSE using a two band tight-binding Hamiltonian
of the semiconductor single electron states and a variational
method to find the exciton binding energy.\cite{mares} The
effective masses of the electron ($m_e^* = 0.067 m_0$) and the
heavy hole ($m_h^* = 0.34 m_0$) determine the tight-binding
hopping parameters $t_e = 1.787 \ eV$ and $t_h = 0.35 \ eV$. The
confining potentials are calculated using a bandgap of $E_g = 1.43
\ eV$ and an offset ratio of $\Delta E_c / \Delta E_v = 67/33$
between the conduction and the valence band. This model, evaluated
on a discretized lattice with 100 sites, reproduces the single
particle energies and wave functions $\Psi_e(x_e)$ and
$\Psi_h(x_h)$ of Ref. \cite{mares} to an accuracy of 1\%.
Following Refs. \cite{mares} and \cite{chuang}, a variational
ansatz for the 1S exciton wave functions, $\Psi_{ex}(x_e,x_h,\rho)
= \sqrt{2/\pi } \Psi_e(x_e) \Psi_h(x_h) \exp{(-\rho / \lambda
)}/\lambda$, is used to minimize their binding energies. Here,
$\rho$ denotes the separation between the electron and the hole in
the plane of the quantum well and perpendicular to the applied
field $ F $, $x_e$ and $x_h$ are the coordinates of the electron
and the hole perpendicular to the plane of the quantum well, and
$\lambda $ is the variational parameter. This wave function is
optimized by minimization of the exciton energy.  The exciton
contribution to the photon absorption spectrum, governed by the
spatial overlap of the electron and hole wave functions, is then
calculated. \cite{chuang} The contribution of the particle-hole
continuum is included to account for the complete absorption
spectrum at zero temperature.

Our approach reproduces the main field-dependent spectral features
of other, more detailed models\cite{mares,chao} of the QCSE in a
simple rectangular potential well profile.  In Fig. 1, we show the
calculated absorption spectra, $\alpha $, as a function of photon
energy $E$ for different electric fields, $F$, applied along the
$x$-direction. As seen in the left panel, an increase of the field
strength leads to a tilt of the confining potential. Consequently,
the electron wave function is shifted towards the right, whereas
the hole wave function moves to the left, resulting in in a
reduced spatial overlap. This in turn causes (i) a shift towards
lower energy, and (ii) a strong reduction in spectral weight of
the dominant excitonic contributions to the photon absorption
spectrum.
 In the lower right panel, the calculated peak exciton
absorption is shown as a function of photon energy for applied
electric fields in the range 0 kV/cm $< F <$ 140 kV/cm.  The arrow
indicates the direction of increasing applied electric field.
Shown as light gray curves are individual spectra for 10 kV/cm
field increments used to calculate the exciton peak absorption
curve. This exciton absorption curve captures the essential
$functionality$ of the QCSE. Maximum exciton absorption decreases
and shifts to lower photon energy with increasing applied electric
field. This rapid loss of resonant behavior in the presence of a
bias voltage dramatically limits the tunability of quantum well
based optical devices.

\begin{figure}[h]
\includegraphics[width=8.5cm]{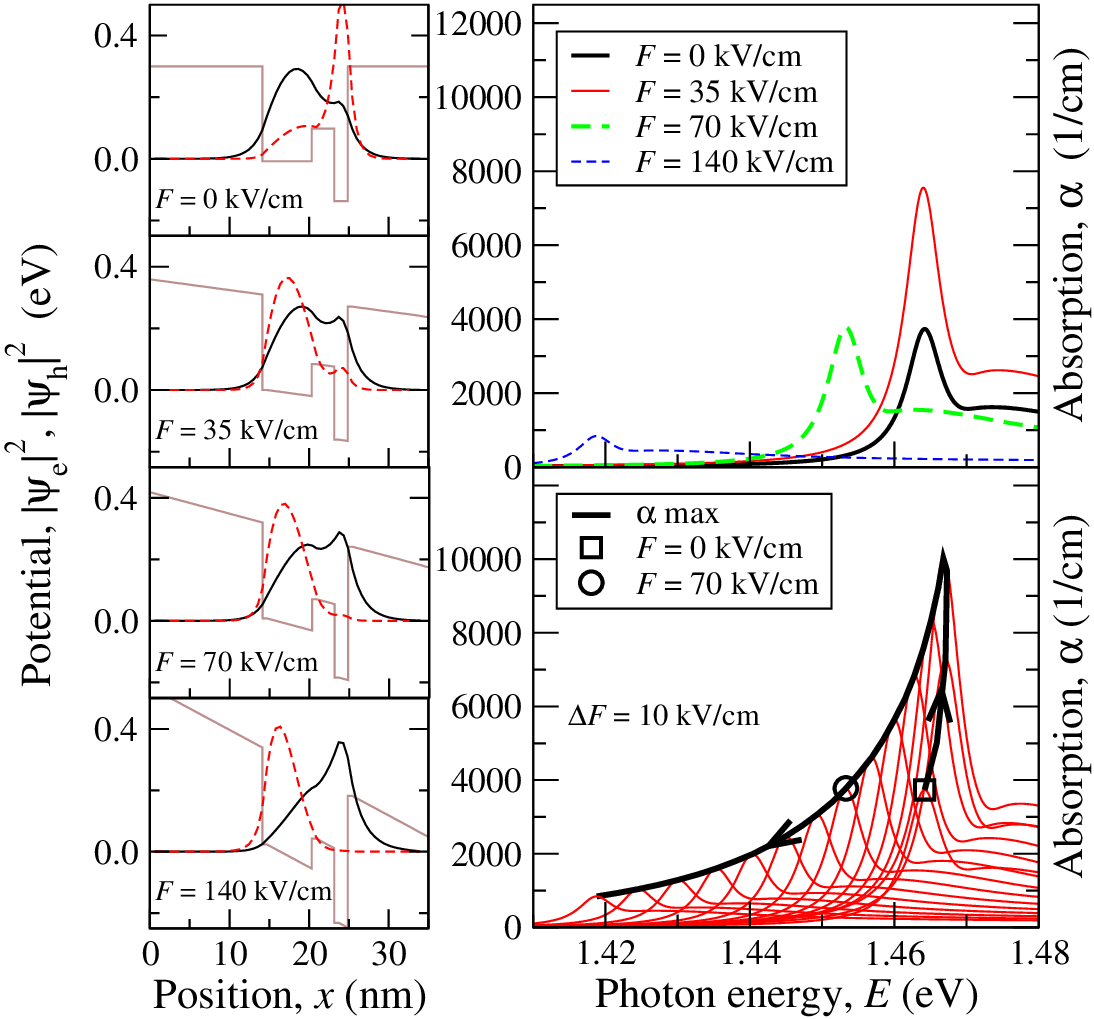}
\caption{ Broken-symmetry double quantum well obtained from
numerical optimization of well width and depth parameters. The
target response is an absorption spectrum with maximized,
equal-height excitonic peaks at bias voltages $ F $ = 0 kV/cm and
$ F $ = 70 kV/cm which are separated in energy by 10 meV. }
\end{figure}

What we wish to show here is that it is possible to use an
adaptive design methodology to create new types of functionality.
For example, consider the situation where we wish to design a
device in which the excitonic absorption peak at $F$ = 0 kV/cm and
$F$ = 70 kV/cm is the same, only shifted in photon energy by at
least 10 meV. This particular functionality would allow us to
rapidly switch the frequency of a quantum well exciton absorption
resonance without loss of its absorption strength. Let us
initially constrain the search for an enabling broken-symmetry
structure to double wells with variable depths and widths. Our
numerical optimization uses a genetic algorithm\cite{whitley} with
a fitness function that simultaneously optimizes the heights and
separation of the exciton peaks at zero and finite (70 kV/cm)
bias. The best solution found by our adaptive quantum design
method for this restricted search is shown in Fig 2. It is
reminiscent of structures investigated
previously.\cite{trezza,susa} The optimized double well causes the
ground state wave function of the hole (broken curve) to develop
two maxima whose relative weight is shifted from left to right as
the electric field is increased. Simultaneously, the center of the
electron wave function (solid curve) moves from left to right,
having a maximum spatial overlap with the right peak of the hole
wave function  at $F$ = 0 kV/cm and with the left peak at $F$ = 70
kV/cm. The resulting exciton peaks in the absorption spectrum,
shown in right panel, have the desired strength and separation.
They are located on two sides of a maximum resonance that is
reached at $F$ = 20 kV/cm. In this broken symmetry structure, the
maximum of the excitonic absorption peak ($\alpha$ max) initially
increases with applied field, and then drops. The corresponding
shift of the resonant energy is also non-monotonic.

\begin{figure}[h]
\includegraphics[width=8.5cm]{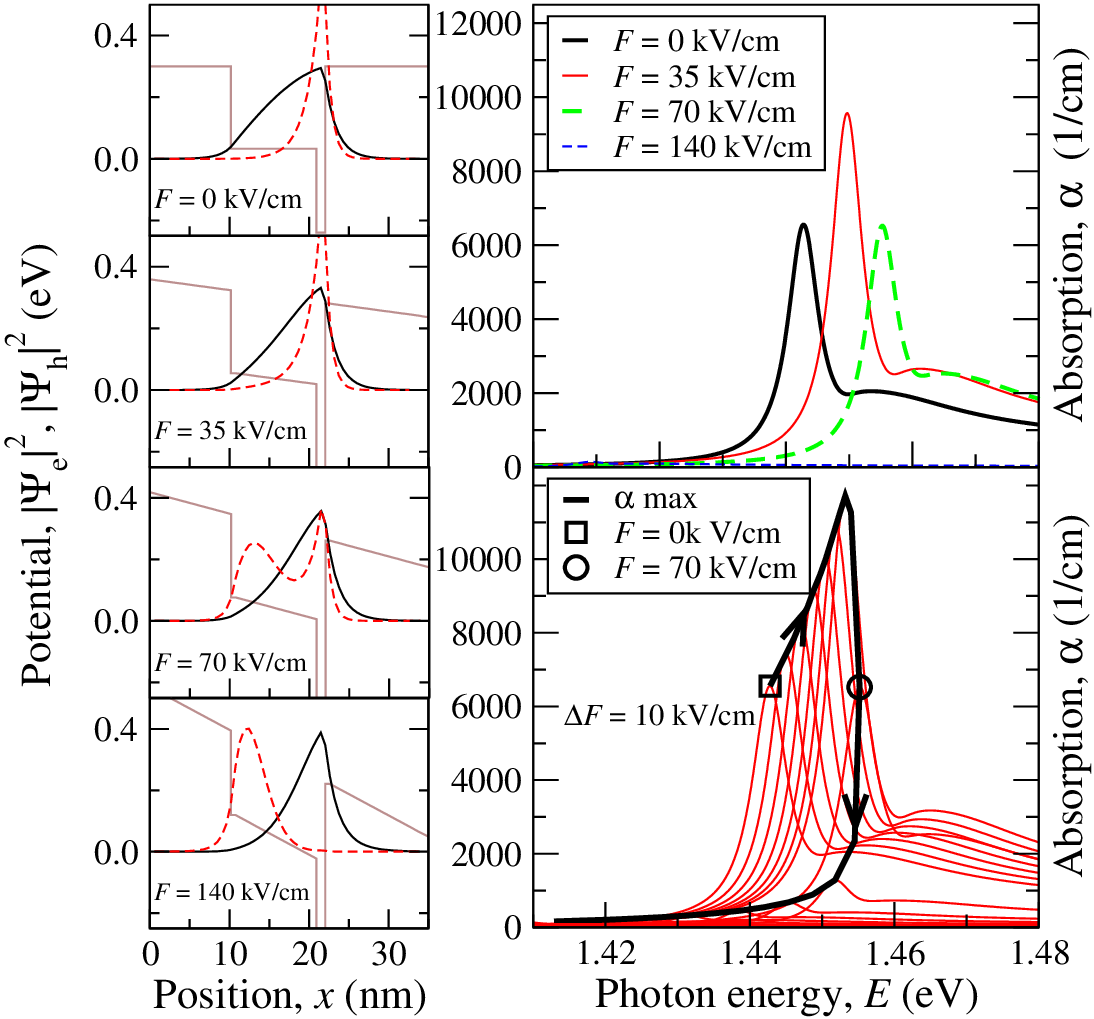}
\caption{ Broken-symmetry quantum well structure obtained from
numerical optimization with additional search parameters compared
to Fig. 2. The target response is an absorption spectrum with
maximized, equal-height excitonic peaks at bias voltages $ F $ =
 0 kV/cm and $ F $ = 70 kV/cm which are separated in energy
by 10 meV. }
\end{figure}

However, there are other, even better solutions if the arbitrarily
imposed initial constraints on the numerical search are relaxed.
In Fig. 3, we show the result of a numerical optimization of the
quantum well in which the positions of the corners of the double
wells are allowed to overlap. The obtained potential profile is
surprisingly simple. The steep drop of $V(x)$ close to the right
boundary of the well pins the hole wave function (broken curve),
whereas the electron wave function (solid curve) is still able to
shift its weight with increasing applied electric field.  The
resulting exciton resonances in $\alpha (E)$ at $F$ = 0 kV/cm and
70 kV/cm are more pronounced than in the solution of Fig. 2.
Furthermore, in this design the exciton absorption curve peak
($\alpha$ max) exhibits a functionality that is different from
that of Fig. 2. It first shifts towards higher energies with
increasing applied electric field before it drops rapidly and
eventually shifts back towards lower photon frequencies. Note,
that this almost vertical drop of $\alpha (E)$ in a very narrow
range of bias voltages indicates an exponential sensitivity that
is highly desirable for the design of quantum well based
modulators.

In summary, the adaptive quantum design methodology used in this
work reveals design options that have not been explored using
conventional ad-hoc approaches. The large number of possible
solutions along with their exponential sensitivity to small
parameter changes render the problem prohibitive to searches by
hand.\cite{susa,chen} Instead, numerical searches identify
optimized broken-symmetry structures which enable desired target
functionalities that are useful in the design of quantum well
based photonic switches and modulators. This new approach allows
us to engineer many-body excitonic wave functions and thus
illustrates a new paradigm in nano-scale design.

We are grateful to Peter Littlewood for discussions. We
acknowledge financial support by DARPA and the Department of
Energy, Grant No. DE-FG03-01ER45908. Computational support was
provided by the USC Center for High Performance Computing and
Communications and by the NERSCC.

\end{document}